\documentclass[10pt,aps,prd,twocolumn,showpacs,nofootinbib,amsmath,amssymb]{revtex4-2}
\usepackage[utf8]{inputenc}

\usepackage{epsf}
\usepackage[dvipsnames]{xcolor}
\usepackage[colorlinks]{hyperref}
\hypersetup{
	colorlinks=true, 
	linktoc=all,     
	linkcolor=blue,  
	citecolor=magenta,
	urlcolor=blue	
}

\usepackage{dcolumn}
\usepackage{bm}

\usepackage{dcolumn}
\usepackage{bm}
\everymath{\displaystyle}

\DeclareMathVersion{monobold}
\SetSymbolFont{letters}{monobold}{OML}{wcrum}{b}{n}
\SetSymbolFont{operators}{monobold}{OT1}{wcrr}{b}{n}
\SetSymbolFont{symbols}{monobold}{OMS}{wcry}{b}{n}

\DeclareMathSymbol{\singleGlyph}{\mathord}{symbols}{5}

\usepackage{cmap}

\usepackage{amssymb}
\usepackage{amsmath}
\usepackage{bm}
\usepackage{graphicx}
\usepackage{mathtools}
\newcommand{\matr}[1]{\mathbf{#1}}

	\def\exp{\mathop{\rm exp}\nolimits}

	\def\exp{\mathop{\rm exp}\nolimits}

	\def\exp{\mathop{\rm exp}\nolimits}
	
\begin{document}

	
\title{Spin of protons in NICA and PTR storage rings as an axion antenna}

\author{\firstname{Nikolai N.}~\surname{Nikolaev}}
\email{nikolaev@itp.ac.ru}
\affiliation{Landau Institute for Theoretical Physics, 	Russian Academy of Sciences,
Moscow Region, Chernogolovka 142432, Russia\\
Moscow Institute of Physics and Technology, Department
of Problems of Theoretical Physics, \\
Moscow region, 141707 Dolgoprudny, Russia}

\begin {abstract}
We discuss a new approach to search for axions in the storage ring experiments, applicable at short coherence time of the in-plane polarization as is the case for protons. The technique can readily be applied at any storage ring equipped with internal polarimeter for the radial polarization of the beam (COSY, NICA, PTR).  We point out a possibility of substantial broadening of the range of attianable axion masses in storage rings with the hybrid electric and magnetic bending, the PTR proton storage ring being an example.
\end{abstract}


\maketitle



\section*{\bf Introduction}

There is a growing interest in the spin of particles  as a highly sensitive spin resonance detector of cosmic axions and axion-like particles. Axions, first proposed by Peccei and Quinn  in 1977  [1]
as a solution to strong CP-violation in QCD,  are widely discussed as a  plausible candidate for the dark matter. One of manifestations of the cold galactic halo axions wil be a NMR-like resonant rotation of the spin in the oscillationg axion field  [2-6].
Here the spin serves as an axion antenna and the experimental search  by the JEDI coillaboration of the axion signal with polarized deuterons in the storage ring COSY is in progress [7].
Since the axion mass is unknown, one is bound to spin precession frequency scanning. In the JEDI experiment one looks for a buildup of the vertical polarization from the idly precessing in-plane polarization when the resonance condition is met during a scan. 

Inherent to the JEDI technique is a need for a long coherence time of the in-plane precessing spin. For instance, it can not be easily extended to protons with arguably short spin coherence time. Besides that, the axion signal depends on the unknown difference of phases of the axion field and idle spin precession. In this communication we suggest the alternative scheme which is free of these limitations and  looks preferred 
one for searches for axions at the Nuclotron, NICA [8] and PTR [9] storage rings. All one needs is an internal  polarimetry which both the  Nuclotron and NICA proper will be  equipped with, and there is no need in the radofrequency spin rotators.  We also comment on exclusive features of the spin frequency scan  in hybrid rings with concurrent electric and magnetic bending, the PRT stotrage ring proposed by the CPEDM collaboration being a good example.

A comprehensive introduction into the physics of axions is found in reviews [4,10-12],
here we only mention the principal points of importance to our discussion, alongside the basics of the spin precession in storage rings [10-15].
When the angular frequency of the axion field is about three times the expansion rate of the Universe, coherent oscillations of the cold cosmological axion field do establish [16-18]. 
Attributing the local energy density of the dark matter $\rho_{\rm DM} \approx 400$ MeV/cm$^3$ [19]
to axions in the invisible halo of our Galaxy, the amplitude of the classical axion field $a(x) = a_0\cos\left(\omega_{a} t- {\bm k}_{a} \cdot {\bm x}\right) $ can be evaluated as [2,20] 
\begin{equation}
a_0 = \frac{1}{m_{a}}\,\sqrt{\frac{2\rho_{\rm DM}\hbar}{c^3}}\,. \label{AxionAmplitude}
\end{equation}
Of key importance is Weinberg's gradient interaction of axions with fermions [21],
\begin{equation} 
L_{a\bar{\psi}\psi} = -\,{\frac{1}{2f_{a}}}\,g_f
\,\overline{\psi}\gamma^\mu\gamma_5\psi\,\partial_{\mu} a(x)\,, \label{AxionFermion}
\end{equation}
which can be reinterpreted as an interaction of fermion's spin with the pseudomagnetic field [22].
The dimensionless constant $g_f \sim 1$ depends on the specific model, and Weinberg derived the relation,  
\begin{equation}
m_{a} \approx m_\pi \frac{f_\pi}{f_{a}}\frac{\sqrt{m_u m_d}}{m_u+m_d}\,,\label{AxionMass}
\end{equation}
where $m_\pi$ and $f_\pi$ are the pion mass and decay constant, and $m_{u,d}$ are mases of light quarks. Still another manifestation of the galactic axion field is the
oscillating contribution to the electric dipole moment (EDM) of nucleons [2,23],
\begin{equation}
\begin{split}
d_N^{\rm ax}(x) =  {\frac{a(x)}{f_a}}\cdot {\frac{\mu_N}{c}}\kappa_{a} \,, \label{OscillatingAxionEDM}
\end{split}
\end{equation}
where $\mu_N$ is the nuclear magneton and the small factor 
\begin{equation}
\begin{split}
\kappa_{(a)} \sim \frac{m_d m_u}{\Lambda_{QCD} (m_d+m_u)}\approx 10^{-2} \label{ChiralSuppression}
\end{split}
\end{equation}
describes the chiral suppression of the EDM  by small masses of light quarks [24,25].

We consider storage rings with crossed magnetic and electric bending fields. The cyclotron angular velocity equals
\begin{equation}\label{Cyclotron}
\bm{\Omega}_c = {\frac {q}{m\gamma}}\left( -{\,\bm B}
+ \frac{{\bm v}\times {\bm E}}{v^2}\right)\, . 
\end{equation}
The contributions to the spin rotation with respect to the particle momentum, $\bm{\Omega}_s = \bm{\Omega}_s^{\rm mdm} + \bm{\Omega}_s^{\rm edm}$,  due to the particle's magnetic dipole moment (MDM), $\mu = g\mu_N$, and the electric dipole moment (EDM), $d$, is given by the generalized BMT equations [15-17]:
\begin{eqnarray}\bm{\Omega}_s^{\rm mdm} &=& {\frac{q}{m}}\left\{-\,G\,\bm{B} + \left(G - {\frac{1}{\gamma^2-1}}
\right){\frac {\bm{v}\times\bm{E}}{c^2}}\right\}\,,\label{Osmdm} \label{MDM}\\
\bm{\Omega}_s^{\rm edm} &=& -d\{ \bm{E} + \bm{v}\times\bm{B}
\}\,.
{}\label{Fukuyama1}
\end{eqnarray}
Here $G= (g-2)/2$ is the magnetic anomaly, ${\bm v}$ and $\gamma$ are particle's velocity and the relativistic $\gamma$-factor, and  the spin tune, $\nu_s$, is defined  via ${\bm \Omega}_s^{\rm mdm}  = \nu_s{\bm \Omega}_c$. 

\section*{\bf Axion effects in all magnetic rings} 
We consider first all magnetic storage rings, ${\bm E}=0$, when $\nu_s = G\gamma$. The salient features of  spin as an axion antenna are as follows. First of all, in the comoving frame (cmf) the axion field induced  EDM interacts with the motional electric field 
\begin{equation}
 {\bm E}_{cmf} = [{\bm v}\times {\bm B}] = -(m\gamma/q)\cdot [\vec{v}\times {\bm \Omega}_c]\, .
 \end{equation} 
 As a spin rotator, it is tantamount to the radiofrequency Wien filter [26] 
 with the radial magnetic field, operated at the axion field angular velocity 
 \begin{equation}
 \omega_{a} = \frac{m_a c^2}{\hbar}\,.
 \end{equation}
 Secondly, spin interacts with the oscillating  pseudomagnetic field, proportional to the particle velocity and  the time derivative $\partial_t a(x)$ [22],
 tangential to the particle orbit and acting as a radiofreqency solenoid. Finally, the velocity of particles in a storage ring, $v$, is of the order of the velocity of light and 
$v\approx 10^3  v_a$, where $v_a \sim 250$ km/s is the velocity of Earth's motion in Galaxy. Hence the pseudomagnetic field acting on the spin of stored particles is three orders in magnitude stronger compared to that acting on, for instance, the spin of ultracold neutrons 
[27,28].

At the axion resonance $\omega_a = \nu_s\Omega_c$, and the instantaneous angular velocity of the axion-driven resonant spin rotation,  derived by Silenko [28],
takes the form 
\begin{eqnarray}
\begin{aligned}
	\bm{\Omega}_{\rm res} = \frac{a_0}{f_{a}}\,\Bigl(&g_f\omega_a\sin(\omega_a t){\frac{\bm{v}}{c}}\\
	& - \,\kappa_a \gamma \cos(\omega_a t)\Bigl[\frac{\bm v}{c}\times\bm{\Omega}_c\Bigr]\Bigr)\,.\label{AxionOmega}
	\end{aligned}
	\end{eqnarray}
The phases of two spin rotators differ by $\pi/2$, but they do rotate spin about orthogonal axes.	Hence upon solving the generalized BMT equations by the Bogoliubov-Krylov averaging [29,30],
one finds the angular velocity of the resonant up-down rotation of the spin envelope
\begin{equation}
\Omega_{res} = {\frac{a_0}{2f_{a}}}\,{\frac{\gamma v}{c}}\, \,|\,g_f G - \kappa_{a}\,|\,\Omega_c\,. \label{OmegaRes}
\end{equation}
Note a strong enhancement of the contribution from the pseudomagnetic field by the factor $g_fG/\kappa_a \gg 1$ compared to the contribution from the axion driven  EDM. In the early simulations of  operation of the storage-ring spin antenna as an axion detector, the contribution of this pseudomagnetic field was not taken into account in [7]
and the actual sensitivity of the JEDI experiment was greatly underestimated.

\section*{\bf Frequency scan search for the axion signal} From now on we focus on the case of the initial vertical polarization $P_{y}$.
At a fixed $\omega_a$, we parameterize a variation of the spin precession angular velocity as $\Omega_s (t) = \omega_a + 2\Omega_t^2 t$, where $2\Omega_t^2 = d\Omega_s/dt$.  Our convention for time is such that a scan starts at negative time $t=-t_0$ and the exact resonance takes place at  $t=0$.  The spin phase during the scan varies as 
\begin{equation}
\theta_s(t) =\omega_a t + \Omega_t^2 t^2
\end{equation}
and duration of the scan is such that $\Omega_t^2 t_0^2 \gg 2\pi$.
A sensitivity to weak axion signal depends on the relationship between the spin coherence time, $\tau_{SCT} = 1/\Gamma$, and the speed of the spin precession frequency scan. 
Specifically, it is easy to derive the time dependence of the axion-driven in-plane polarization envelope $P_{xz}$:
\begin{equation}
P_{xz}(t)=P_y \Omega_{res} \int_{-t_0}^{t}d\tau \exp\left[-\Gamma(t-\tau)\right]\cos(\Omega_t^2 \tau^2)\, . \label{Master}
\end{equation}

In the usually discussed scheme with the in-plane initial polarization [6,7,31,32],
the axion signal will be proportional to $\sin \Delta$ [29,33],
where $\Delta$ is an entirely unknon difference of the spin presession and axion oscillation phases.  As a practical remedy, it was proposed to fill the ring with four bunches of different polarizations [6,7]. 
In contrast to that, a buildup of the in-plane polarization from the initial vertical one is free of this phase ambiguity.

Numerical evaluation of the integral (\ref{Master}) is not a problem, here we proceed to simple analytic evaluations, leaving aside the issue of numerical simulations. First we start with the limit of large spin coherence time,
$\Gamma \ll \Omega_t$, when the exponential damping factor in the integrand can be neglected, what is the case for stored deuterons. Then the envelope $P_{xz}$ will exhibit a jump of temporal width $t_1 \sim 1/\Omega_t$ and the amplitude 
$P_{xz}^{max} \sim P_y\Omega_{res}/\Omega_t$. The observed signal will be a jump in the up- down asymmetry in the polarimeter oscillating with the spin precession frequency. A Fourier analysis technique for extraction of the spin envelope form such an asymmetry data has been developed by the JEDI collaboration [34,35].

The opposite limiting case of $\Gamma \gg \Omega_t$ is of interest for stored protons which, because of the large magnetic anomaly, will arguably have a short spin coherence time [34].
 Here emerges a a new time scale
\begin{equation}t_2 \sim \frac{\Gamma}{\Omega_t^2}\, . 
\end{equation}  
Algebraically at times $-t_2 <t< t_2$ the envelope $P_{xz}$ will oscillate with  constant amplitude,
\begin{equation}
P_{x,z} \approx \frac{\Omega_{res}}{\Gamma}  P_y\cos(\Omega_t^2 \tau^2)\, , \label{Plateau}
\end{equation}
while oscillatios will vanish beyond this interval. The practical Fourier analysis of the up-down asymmetry from the in-plane precessing polarization returns the envelope  as  a positive defined quanity, and the axion signal will look like a corrugated  plateau of the height $ |P_{xz}|\sim \Omega_{res}/\Gamma$ and width $t_2$. 

A lesson from  searches for resonances is that at similar statistics broad resonances are much harder to be identified. However, in our case a salient feature of the signal (\ref{Plateau}) is a central peak of width $\sim 1/\Omega_t$, the peaks at wings will have a spacing  $\sim \tau_{SCT}$ and can be resolved if sufficient statistics can be accumulated during the spin coherence time. Fits to the well specified function (\ref{Plateau}) will facilitate inentification of the axion signal. Furthermore, the central peak will be better visible at faster ramping of the cyclotron frequency such that  
\begin{equation}
\frac{d\Omega_{c}}{dt} \sim \Gamma^2\,.
\end{equation}

Still another option is to run the proton ring at magic energies with strong suppression of the spin decoherence as was suggested in [36]. These magic  energies are roots of the equation
 \begin{equation}
\begin{aligned}
1- \frac{c^2}{v^2} \cdot\Bigl(1+\frac{K}{G\gamma}\Bigr)\cdot\Bigl(\frac{1}{\gamma_{tr}^2}-\frac{1}{\gamma^2}\Bigr)=0\,,\quad K= \pm 1, \pm 2,...
\end{aligned}
\end{equation}
exact position of magic energies depends on the transition $\gamma$-factor, $\gamma_{tr}$, of the particular ring. For instance, at COSY the roots exist at negative $K$. The lowest magic 
kinetic energy for protons is $T_p \approx 30 $ MeV at $K=-2$, and as an axion antenna it will be tuned to search for axions with
$\omega_a =(2-G_p)\Omega_c$, where $G_p=1.793$ is the magnetic anomaly of the proton. The next magic energy at $K=-3$ is $T_p \approx 130$ MeV, and COSY the will be tuned to $\omega_a = (3-G_p)\Omega_c$. At still higher magic energy $T_p \approx 210$ MeV one will probe $\omega_a= (4-G_p)\Omega_c$. Note that $\Omega_c$ scales with $\propto \gamma/\sqrt{\gamma^2 -1}$.

\section*{\bf Axion effects in hybrid storage rings} 
Note that in all magnetic fields the resonance condition entails $\omega_a= p/(Rm)$, where $R$ is the ring radius and $p$ is the particle momentum. Consequently, as an axion antenna the attainable axion masses are bounded from below by the minimal momentum the storage ring can run at.
The hybrid ring with concurrent magnetic and electric bendings are much more versatile compared to all magnetic rings.  An example is provided by the prototype test ring PTR proposed by the CPEDM collaboration [9].
With the radial electric field $E_0 = 7\times 10^6$ V/m  complemented by the vertical magnetic field  $B_0 = 0.0327\ {\rm T}$,  PTR will provide the frozen spin of protons,
${\bm \Omega}_s^{\text mdm}=0$, {it i.e., $\nu_s = 0.$}. Beyond this point, the electric and magnetic fields must be varied synchronously to  preserve the injection energy, the orbit radius and the cyclotron frequency,
\begin{equation}
\Delta {\bm B} = \frac{1}{v^2}[{\bm v}\times \Delta {\bm E}]\, ,
\end{equation}
see Eq. (\ref{Cyclotron}). Then according to Eq. (\ref{MDM}) the spin tune will vary  as 
\begin{equation}
\nu_s= -G_p\gamma \frac{\Delta E}{E_0}\,,
\end{equation}
taking the canonical value, $\nu_s =G\gamma$, in the all-magnetic limit of $\Delta E =-E_0$. In the same range the magnetic field must be raised concurrently to 
\begin{equation}
B = \frac{G_p+1}{1-G_p \gamma^2 \beta^2} B_0 = 3.39 B_0\,. 
\end{equation}
The practically attainable range of magnetic fields will depend on the design of the air coil. We emphasize that arguably the spin coherence time $\tau_{SCT}$ scales with $1/\nu_s^2$ [36], which specifically favors operation of the  hybrid PTR as an axion antenna to search for small mass axions.

Note that at the fixed orbit of the beam the electric field in the comoving frame, acting on proton's EDM, does not depend on $\Delta E$, while the axion resonance will take place at
\begin{equation}
\omega_a = -G_p\gamma \Omega_c\frac{\Delta E}{E_0}\,. \label{ScanBand}
\end{equation}
The resulting angular velocity of the axion-driven spin rotation will be given by 
\begin{equation}
\Omega_{res} = {\frac{a_0}{2f_{(a)}}}\,{\frac{\gamma v}{c}}\, |\,g_f G \frac{\Delta E}{E_0}+ \kappa_{(a)}\,|\,\Omega_c\,,  \label{OmegaHybrid}
\end{equation}
with the pseudomagnetic field contribution suppressed by the factor $\Delta E/E_0$. Increasing the electic field in delectors must be viewed impractical.

A hybrig storage ring operating as a broadband axion antenna at a fixed beam energy has several advantages. We mention injection at a fixed energy, running the radiofrequency cavity in a narrow range of frequencies, a polarimetry optimized for one fixed energy.   

\section*{\bf Summary} We demonstrated how the spin of polarized protons can be used as an axion antenna in spite of the short spin coherence time of the in-plane polarization of protons. The key point is to look for a buildup of the idly precessing in-plane polarization starting from the stable verical one.  Of special interest is a possibility to enhance a sensitivity to axions working at selected beam energies with strong suppression of spin decoherence effects. Hybrid prototype rings emerge as a promising proton spin
antenna sensitive to axions with  small masses up to few neV/c$^2$. These observations suggest new options for  the experimental searches for axions at NICA in Dubna, COSY in Juelich, planned PTR and elsewhere. 
We emphasize that our approach can readily be extended to the hydrid deuteron storage rings.

\section*{\bf Acknowledgements} A support of thus study by the Russian Science Foundation  grant 22-42-04419 is acknolwledged. We are grateful to A.J. Silenko for useful discussions.

\section*{\bf References}

[1] R. D. Peccei and H. R. Quinn, Phys. Rev. Lett. {\bf 38}, 1440
(1977).

[2] P. W. Graham and S. Rajendran, Phys. Rev. {\bf D 88},
035023 (2013).

[3] D. Budker, P. W. Graham, M. Ledbetter, S. Rajendran,
and A. Sushkov, Phys. Rev. {\bf X 4}, 021030 (2014).

[4] P. Sikivie, Rev. Mod. Phys. {\bf 93}, 015004 (2021).

[5] C. Abel, N.J. Ayres, G. Ban,  et al., Phys. Rev. {\bf  X 7}, 041034 (2017).

[6] S. P. Chang, S. Hacıomeroglu, O. Kim, et al., Phys. Rev. {\bf D 99}, 083002 (2019).

[7] J. Pretz, S. Karanth, E. Stephenson, et al., Eur.
Phys. J. {\bf C 80}, 107 (2020).

[8] N. N. Agapov, V. D. Kekelidze, A. D. Kovalenko, et al., 
Physics-Uspekhi {\bf 59}, 583 (2016)

[9] F. Abusaif, A.  Aggarwal, A.  Aksentev, et al. (CPEDM Collaboration), Storage ring
to search for electric dipole moments of charged particles:
Feasibility study, CERN Yellow Reports: Monographs,
2021-003 (CERN, Geneva, 2021).

[10] L. Di Luzio, M. Giannotti, E. Nardi, and L. Visinelli,
Phys. Rept. 870, {\bf 1} (2020).

[11] C.B. Adams, A. Agrawal, R. Balafendiev, et al., in 2022 Snowmass Summer Study
(2022) arXiv:2203.14923.

[12] S. N. Vergeles, N. N. Nikolaev, Y. N. Obukhov,
et al., (2022), 
arXiv:2204.00427 [hep-th].

[13 V. Bargmann, L. Michel, and V. L. Telegdi, Phys. Rev.
Lett. {\bf 2}, 435 (1959).

[14] D. F. Nelson, A. A. Schupp, R. W. Pidd, and H. R.
Crane, Phys. Rev. Lett. {\bf 2}, 492 (1959).

[15] Fukuyama and A. J. Silenko, Int. J. Mod. Phys. {\bf A 28},
1350147 (2013).

[16] J. Preskill, M. B. Wise, and F. Wilczek, Phys. Lett. {\bf  B
120}, 127 (1983).

[17] L. F. Abbott and P. Sikivie, Phys. Lett. {\bf B 120}, 133
(1983).

[18] M. Dine and W. Fischler, Phys. Lett. {\bf B 120}, 137 (1983).

[19] J. Read, J. Phys. G: Nucl. Part. Phys. {\bf 41}, 063101 (2014).

[20] P. Sikivie, Lect. Notes Phys. {\bf 741}, 19 (2008), Springer Verlag (Berlin, Heidelberg),

[21] S. Weinberg, Phys. Rev. Lett. {\bf 40}, 223 (1978).

[22] M. Pospelov, A. Ritz, and M. B. Voloshin, Phys. Rev.
{\bf D 78}, 115012 (2008).

[23] P. W. Graham and S. Rajendran, Phys. Rev. {\bf D 84},
055013 (2011).

[24] V. Baluni, Phys. Rev. {\bf D 19}, 2227 (1979).

[25] R. J. Crewther, P. Di Vecchia, G. Veneziano, and E. Witten,
Phys. Lett. {\bf B 88}, 123 (1979), [Erratum: Phys.Lett. {\bf B
91}, 487 (1980)].

[26] J. Slim, R. Gebel, D. Heberling, et al., Nucl. Instr. Meth. Phys. Res. {\bf A 828}, 116
(2016).

[27] P. W. Graham, S. Haciomeroglu, D. E. Kaplan, et al., Phys.
Rev. D 103, 055010 (2021), arXiv:2005.11867.

[28] A. J. Silenko (2021), arXiv:2109.05576.

[29] A. Saleev, N. N. Nikolaev, F. Rathmann, et al. (JEDI), Phys. Rev. Accel. Beams {\bf 20},
072801 (2017).

[30] A. J. Silenko, EPL {\bf 118}, 61003 (2017).

[31] E. Stephenson, PoS PSTP {\bf 2019}, 018 (2020).

[32] S. Karanth, “New method to search for axion-like particles
demonstrated with polarized beam at the cosy storage
ring,” (2021), dPG Spring Meeting: Dortmund, 15-
19 March 2021.

[33] F. Rathmann, N. N. Nikolaev, and J. Slim, Phys. Rev.
Accel. Beams {\bf 23}, 024601 (2020), arXiv:1908.00350.

[34] Z. Bagdasarian, S. Bertelli, D. Chiladze, et al., Phys. Rev. ST Accel. Beams {\bf 17},
052803 (2014).

[35] D. Eversmann, V. Hejny, F. Hinder, et al. (JEDI), Phys. Rev. Lett. {\bf 115},
094801 (2015).

[36] A. Lehrach, B. Lorentz, W. Morse, N. Nikolaev, and
F. Rathmann, (2012), arXiv:1201.5773 [hep-ex].

\end{document}